\RequirePackage{ifpdf}
\documentclass[12pt,letterpaper]{JHEP3}
\pdfoutput=1

\usepackage{graphicx}                     
\usepackage{amsmath}
\usepackage{amsfonts}
\usepackage{amsthm}
\usepackage{latexsym}                  
\title{$S^2\times S^3$ geometries in ABJM and giant gravitons}
\author{Yolanda Lozano \footnote{\texttt{ylozano@uniovi.es}}$^{\hspace{-0.15cm}1}\hspace{0.1cm} $, 
\hspace{-0.1cm} Andrea Prinsloo \footnote{\texttt{a.prinsloo@surrey.ac.uk}}$^{\hspace{-0.15cm}2}\hspace{0.1cm} $ \\
~ \\
\hspace{-0.6cm} $^{1}$ Departamento de F{\'\i}sica, Universidad de Oviedo,  \\
\hspace{-0.6cm} \hspace{0.165cm} Avda.~Calvo Sotelo 18, 33007 Oviedo, Spain. \\

\hspace{-0.6cm} $^{2}$ Department of Mathematics, University of Surrey, \\
\hspace{-0.6cm} \hspace{0.165cm} Guildford, GU2 7XH, United Kingdom.\\}

\abstract{
We construct a new NS5-brane solution in $AdS_4\times \mathbb{CP}^3$ with $S^2\times S^3$ topology. This solution belongs to the general class of non-Einstein $N_{11}$ metrics to which $T^{1,1}$ belongs, and carries a non-vanishing D0-brane charge. In eleven dimensions it gives rise to a squashed $S^2\times S^3$ M5-brane giant graviton which is now of the $N_{10}$ type.
The energies  of both solutions satisfy the BPS bound $E=k\hspace{0.05cm}Q/2$, indicating supersymmetric configurations, where $Q$ is interpreted as D0-brane charge for the NS5-brane and as angular momentum for the dynamically stable M5-brane giant graviton. The ground state is degenerate with a spherical D2 or M2-brane, rather than with a point-like object. Moreover, while the charge of the spherical 2-brane can be arbitrary, the charge of the $S^2\times S^3$ 5-brane is bounded by $N/2$, with $N$ the rank of the ABJM gauge group, a manifestation of the stringy exclusion principle.  A microscopic description, suitable for the study of the finite 't Hooft coupling region, is provided in terms of spherical D2 or M2-branes expanding into fuzzy 3-spheres.}

\preprint{\small{FPAUO--13/01} \\
\small{DMUS--MP--13/04}}

\begin{document}

\section{Introduction}

Giant gravitons have allowed for substantial progress in developing the precise dictionary between the two sides of the AdS/CFT correspondence \cite{Maldacena:1997re}.  On the gravity side of the correspondence, they are realized as dynamically stable D-brane configurations, which (typically) wrap contractible cycles in the ten dimensional supergravity geometry.  Their extension is supported by a coupling to RR-potentials and by their angular momentum in the compact space \cite{McGreevy:2000cw,Grisaru:2000zn}. In maximally supersymmetric Yang-Mills (MSYM) theory, giant gravitons are described by subdeterminants and, more generally, by Schur polynomials of the adjoint scalars which constitute the Higgs sector of MSYM theory \cite{Balasubramanian:2001nh,Corley:2001zk,de Mello Koch:2007uu,de Mello Koch:2007uv,Bekker:2007ea}.  These Schur polynomial operators realize quite explicitly some of the characteristic properties of the dual D-brane configurations (see for instance \cite{Balasubramanian:2004nb,Berenstein:2006qk}), as well as the stringy exclusion principle \cite{Maldacena:1998bw}. They also encode non-trivial global and local geometric properties, thus providing a concrete realization of geometry as an emergent phenomenon, arising from interactions in a strongly-coupled quantum field theory with a large number of gauge degrees of freedom \cite{Berenstein:2005aa,Berenstein:2006yy,LLM}.

The ABJM correspondence \cite{Aharony:2008ug} relates both M-theory on $AdS_{4} \times S^{7}/\mathbb{Z}_{k}$ and type IIA string theory on $AdS_{4}\times \mathbb{CP}^{3}$ with a Super-Chern-Simons-matter theory known as the ABJM model.  Due to the non-trivial geometry of the orbifold and the complex projective space, this constitutes a particularly interesting framework in which to test these ideas. Giant gravitons realized as M-branes in $AdS_{4}\times S^{7}$ were originally constructed in \cite{Grisaru:2000zn,Mikhailov}, but comparatively little is known about the operators dual to M-brane giant gravitons in the ABJM model (although progress has been made in \cite{Berenstein:2008dc,SheikhJabbari:2009kr,Berenstein:2009sa,Dey,dMMMP,Caputa:2012}). New supersymmetric giant gravitons in the type IIA string theory with quite non-trivial geometric and topological properties were studied in \cite{Nishioka:2008ib,Herrero:2011bk,Giovannoni:2011pn} (see also \cite{Hamilton:2009iv}).  Supersymmetric black ring and giant tori configurations were constructed as AdS giant gravitons on $AdS_4\times S^7/\mathbb{Z}_k$ \cite{Nishioka:2008ib}. The dual BPS operators should not only encode Gauss' law, but also shed light on how a topological invariant such as genus can be seen in the gauge theory. A class of candidate duals has been proposed in \cite{Berenstein:2009sa} (see also \cite{Kim}), although still more work is needed to elucidate many of their properties.  Other types of supersymmetric configurations involving non-trivial tensionless M2-branes were discussed in \cite{Carballo:2009ei}.

The construction of supersymmetric higher dimensional non-trivial geometries in type IIA string theory on $AdS_{4}\times \mathbb{CP}^{3}$ from (genuine) M-brane giant gravitons on $AdS_4\times S^7/\mathbb{Z}_k$ was recently explored in \cite{Herrero:2011bk}. One such example was a twisted 5-sphere supported by D0-brane charge. This configuration exhibits interesting properties not found in simpler geometries.  While its ground state is degenerate with $Q_{D0}$ D0-branes, this finite size NS5-brane has much in common with a giant graviton. In particular, the D0-charge must be bounded by the rank of the gauge group, $N$, thus providing an explicit realization of a stringy exclusion principle, this time in terms of expanded D0-branes. Moreover, in the maximal case, $Q_{D0}=N$, the NS5-brane collapses to a $\mathbb{CP}^2$, and its energy can be accounted for both by a bound state of $k$ D4-branes wrapping the $\mathbb{CP}^2$  (dual to $k$ determinant operators, a special case of  di-baryon operators \cite{GLR,Murugan:2011zd},  with $k$ the level number of the gauge group) and by $N$ D0-branes. This configuration provides an explicit realization in string theory of the symmetry of Young diagrams with $N$ rows and $k$ columns, where a single row gives a D0-brane and a single column a D4-brane \cite{Aharony:2008ug}. It also provides an explicit example of the idea of \cite{Aharony:2008ug} that this instability in the field theory could be realized in the string theory as an NS5-brane instanton turning the $k$ D4-branes into $N$ D0-branes. 

The D4-brane giant graviton of \cite{Giovannoni:2011pn} is another example of a non-trivial supersymmetric higher dimensional geometry in $AdS_{4}\times \mathbb{CP}^{3}$.  This dynamically stable D4-brane is wrapped on a geometrically non-trivial (but contractible) 4-manifold in the $\mathbb{CP}^{3}$ compact space and factorizes at maximum size into two D4-branes wrapping $\mathbb{CP}^{2}$ cycles, each dual to a di-baryon operator built from a bifundamental scalar \cite{GLR,Murugan:2011zd}. The BPS operator dual to this D4-brane giant graviton is a totally antisymmetric Schur polynomial, constructed from an adjoint composite scalar field, which is built from two bifundamental scalars \cite{Dey,dMMMP}. Strong evidence was found of a geometric dependence encoded in the spectrum of small fluctuations about this D4-brane giant graviton, which we anticipate should be visible in the excitation spectrum of the dual operators \cite{Giovannoni:2011pn,Hirano:2012vz}.
        
In this paper we provide further examples of higher dimensional non-trivial geometries in $AdS_4\times \mathbb{CP}^3$ and $AdS_4\times S^7/\mathbb{Z}_k$, which can be used to investigate the emergence of geometry and topology from the ABJM model.  We construct an NS5-brane in type IIA string theory on $AdS_4\times \mathbb{CP}^3$ with an $S^{2}\times S^{3}$ geometry, belonging to the general class of non-Einstein $N_{11}$ type metrics discussed in \cite{Candelas:1989js}\footnote{The $T^{1,1}$ is the only Einstein example in this $N_{11}$ class.}.  The extension of this NS5-brane is supported by D0-brane charge on its worldvolume and by its coupling to the RR potentials, $C_{5}$ and $C_{1}$. It satisfies a BPS bound and is energetically degenerate, in its ground state, with a D2-brane with D0-brane charge wrapping the $S^{2}$.  This NS5-brane also carries a topological charge associated with one-dimensional topological defects interpreted as the end-points of D2-branes ending on the NS5-brane.  While a spherical D2-brane can have arbitrary D0-brane charge, the same is not true of the expanded NS5-brane configuration, which must have D0-brane charge $Q_{D0} \leq N/2$.  This is a realization of a stringy exclusion principle. As for the NS5-brane configuration of \cite{Herrero:2011bk}, in the maximal case, the energy of the NS5-brane can be accounted for both by $N/2$ D0-branes and by $k/2$ D4-branes dual to di-baryons. This is again a manifestation of the symmetry of the Young tableaux with $N$ rows and $k$ columns. In eleven dimensions the NS5-brane is realized as an M5-brane giant graviton with the geometry of a squashed $S^2\times S^3$ of the $N_{10}$ type  \cite{Candelas:1989js}. The ground state is not degenerate with a point-like graviton, as in the usual case \cite{McGreevy:2000cw,Grisaru:2000zn}, but with a M2-brane graviton wrapping an $S^{2}$. This happens because the M5-brane wraps a topologically non-trivial manifold $S^{2}\times S^{3}$ and thus cannot shrink to a point-like object.   This M5-brane giant graviton again carries a topological charge, which we interpret as associated with M2-branes ending on the M5-brane.  The momentum of the M5-brane must be less than $N/2$, again a realization of a stringy exclusion principle. Microscopically the $N_{11}$ and $N_{10}$ 5-branes arise from the expansion of spherical 2-branes into fuzzy 3-spheres ($S^{1}$ bundles over fuzzy 2-spheres) through the Myers dielectric effect \cite{Myers:1999ps}. The topological charge associated with these configurations arises from gravitational waves ending on the spherical 2-branes.

The organization of this paper is as follows: In section 2, we construct the NS5-brane solution. Section 2.1 contains a brief discussion of the $AdS_{4} \times \mathbb{CP}^{3}$ background and of some of the properties of the theory of $U(1)$ bundles over $S^2\times S^2$ manifolds relevant for our later analysis. Then, in section 2.2, we present our ansatz for the NS5-brane solution. In section 2.3, we show that the brane is stabilized by a non-vanishing D0-brane charge which couples to the 5-form potential $C_{5}$. Topological charge associated with D2-branes ending on the NS5-brane must also be included so that the BPS bound $E=k \hspace{0.05cm} Q_{D0}/2$ can be reached. The expanded configuration exhibits interesting properties such as the existence of a bound on the D0-brane charge and a symmetry between di-baryons and di-monopoles that we discuss.  In section 3, we uplift the NS5-brane solution to eleven dimensions and show that it corresponds to an M5-brane propagating on the $S^1/\mathbb{Z}_k$ direction, which now has the geometry of a squashed $S^2\times S^3$ of the $N_{10}$ type. Allowing again for topological charge associated with M2-branes  ending on the M5-brane, we show that the ground state is a giant graviton which is degenerate with a spherical M2-brane. Section 4 contains the microscopical description of the $S^2\times S^3$ 5-branes.
The fact that the NS5/M5-brane configurations are topologically non-trivial implies that the microscopic description, suitable for a study of the finite 't Hooft coupling region, should be given in terms of spherical D2/M2-branes carrying D0-brane charge/angular momentum.  We show that at strong coupling D2-branes (M2-branes) can expand via the Myers dielectric effect into a fuzzy NS5-brane (M5-brane). This 5-brane is topologically an $S^2$ times a fuzzy $S^3$, of the type discussed in \cite{Janssen:2003ri, Nastase:2009ny}. Finally, section 5 contains our conclusions and directions for future research.

\section{The $S^2\times S^3$ NS5-brane}

\subsection{The background}

Let us start by collecting some relevant information about the type IIA ten dimensional supergravity background $AdS_4\times \mathbb{CP}^3$ and its relation with $N_{11}$ manifolds \cite{Candelas:1989js}.

\subsubsection{Type IIA 10D supergravity background $AdS_4\times \mathbb{CP}^3$}

We describe the $AdS_4\times \mathbb{CP}^3$ background in global coordinates for the $AdS_4$ part: 
\begin{eqnarray}
\label{theAdS}
ds^2 &=& \frac{L^{2}}{4}  \,  ds^2_{AdS_4}+L^2 \, ds^2_{\mathbb{CP}^3} \nonumber\\
& =&  \frac{L^{2}}{4} \left\{ -\left(1+r^2\right) dt^2+\frac{dr^2}{\left(1+r^2\right)}+r^2 \, d\Omega_2^2\right\} +L^2 \, ds^2_{\mathbb{CP}^3}\, ,
\end{eqnarray}
with $L$ the radius of curvature in string units:
\begin{equation}
L=\left(\frac{32\pi^2 N}{k}\right)^{1/4}\, ,
\end{equation}
and make use of the following parameterization for the $\mathbb{CP}^3$ space \cite{Nilsson:1984bj}:
\begin{eqnarray}
\label{theCP^3}
&& \hspace{-0.25cm} 
ds_{\mathbb{CP}^3}^2  = d\zeta^2+\frac14 \Bigl[\cos^2{\zeta}\sin^2{\zeta}\,\Bigl(d\psi+\cos{\theta}_1 \, d\phi_1+\cos{\theta_2} \, d\phi_2\Bigr)^2+\\
&& \hspace{-0.25cm} \hspace{3.0cm} + \, \cos^2{\zeta}\,\Bigl(d\theta_1^2+\sin^2{\theta}_1 \, d\phi^2_1\Bigr) + \sin^2{\zeta}\,\Bigl(d\theta_2^2+\sin^2{\theta}_2 \, d\phi_2^2\Bigr)\Bigr], \nonumber \hspace{1.0cm}
\end{eqnarray}
where $0\le \zeta < \frac{\pi}{2}$, $0\le \psi < 4\pi$, $0\le \phi_i\le 2\pi$ and $0\le \theta_i < \pi$.
The background is supported by $F_2$ and $F_6$ fluxes which in these coordinates are given by:
\begin{eqnarray}
\label{F2}
F_2&=& -\frac12 \, k \, \Bigl\{\sin{2\zeta}\,d\zeta \wedge [d\psi + \cos{\theta_1}\,d\phi_1+ \cos{\theta_2}\,d\phi_2]+\nonumber\\
&&\hspace{1.2cm} +\cos^2{\zeta}\sin{\theta_1}\,d\theta_1\wedge d\phi_1-\sin^2{\zeta}\sin{\theta_2}\,d\theta_2\wedge d\phi_2\Bigr\}\, ,   \\
F_6&=&\frac18\, k L^4\, \, {\rm dVol}(\mathbb{CP}^3)\, .
\label{F6}
\end{eqnarray}
The dilaton reads:
\begin{equation}
e^\phi=\frac{L}{k}\, .
\end{equation}

\subsubsection{The $\mathbb{CP}^3$ for constant $\zeta$ as an $N_{11}$ manifold}

In the parameterization (\ref{theCP^3}) it is manifest that, for constant $\zeta$, the $\mathbb{CP}^3$ becomes a non-trivial $U(1)$ fibre over $S^2\times S^2$, which is of the $N_{11}$ type discussed in \cite{Candelas:1989js}.
Since this is the ansatz that we will take for the NS5-brane in the next subsection, we now recall some properties of $N_{pq}$ manifolds relevant for our later discussion.  These $N_{pq}$ manifolds form the bases of cones embedded into $\mathbb{C}^{4}$.  The reader is referred to \cite{Candelas:1989js} for more details.

The metrics on the manifolds $N_{pq}$, with $p,q$ relatively prime integers, are fibre bundles over $S^2\times S^2$ with $U(1)$ fibres:
\begin{eqnarray}
\label{Npq}
d\Sigma_{pq}^2&=&\lambda^2\,\Bigl(d\psi+p\cos{\theta}_1\,d\phi_1+q\cos{\theta}_2\,d\phi_2\Bigr)^2 +\nonumber\\
&&+\,\Lambda_1^{-1}\,\Bigl(d\theta_1^2+\sin^2{\theta_1}\,d\phi_1^2\Bigr)+\Lambda_2^{-1}\,\Bigl(d\theta_2^2+\sin^2{\theta_2}\,d\phi_2^2\Bigr)\, .
\end{eqnarray}
If the conditions on the constants $\lambda$, $\Lambda_1$, $\Lambda_2$:
\begin{eqnarray}
4&=&\frac{\lambda^2}{2}[(p\Lambda_1)^2+(q\Lambda_2)^2]\nonumber\\
&=&\Lambda_1-\frac12 (\lambda\, p \Lambda_1)^2\nonumber\\
&=&\Lambda_2-\frac12 (\lambda\, q \Lambda_2)^2
\end{eqnarray}
are fulfilled, then the metrics are Einstein and the corresponding cones Ricci-flat. 

For the two choices $p=q=1$; $p=1$, $q=0$, the fibre bundles are $S^2\times S^3$. These two metrics cannot be mapped onto each other, and 
represent different geometries on $S^2\times S^3$.
Further, for suitable choices of $\lambda$, $\Lambda_1$, $\Lambda_2$ they are Einstein. This corresponds to the $T^{1,1}$ for $p=q=1$:
\begin{equation}
d\Sigma_{11}^2=\frac{1}{9}\Bigl(d\psi+\cos{\theta}_1d\phi_1+\cos{\theta}_2d\phi_2\Bigr)^2+\frac16\Bigl(d\theta_1^2+\sin^2{\theta_1}d\phi_1^2\Bigr)
+\frac16\Bigl(d\theta_2^2+\sin^2{\theta_2}d\phi_2^2\Bigr)
\end{equation}
and to:
\begin{equation}
d\Sigma_{10}^2
=\frac18\Bigl(d\psi+\cos{\theta}_1d\phi_1\Bigr)^2+\frac18\Bigl(d\theta_1^2+\sin^2{\theta_1}d\phi_1^2\Bigr)
+\frac14\Bigl(d\theta_2^2+\sin^2{\theta_2}d\phi_2^2\Bigr)\, ,
\end{equation}
for $p=1$, $q=0$. Clearly, the $\mathbb{CP}^3$ space with constant $\zeta$ belongs to the general class of non-Einstein $N_{11}$ manifolds described by (\ref{Npq}). This connection with $N_{11}$ manifolds was in fact used in  \cite{Giovannoni:2011pn} in order to find the right ansatz for the D4-brane giant graviton on $AdS_4\times \mathbb{CP}^3$ from that of the D3-brane giant graviton on $AdS_{5}\times T^{1,1}$, previously constructed in \cite{Hamilton:2010sv}.

\subsection{The NS5-brane ansatz}

Let us now take an NS5-brane at the origin of the $AdS_{4}$ space and wrapping the non-Einstein $N_{11}$ manifold contained in the $\mathbb{CP}^3$ space at constant $\zeta$. The metric induced on the NS5-brane worldvolume is given by
\begin{equation}
\label{NS5ansatz}
ds^2_{\text{NS5}}= \frac{L^2}{4}\Bigl[ -dt^2 + \cos^2{\zeta}\sin^2{\zeta}\,\Bigl(d\psi+A_1+A_2\Bigr)^2+\cos^2{\zeta}\, \, ds^2_{S_1^2}+\sin^2{\zeta}\,\, ds^2_{S_2^2}\Bigr],
\end{equation}
where $A_i=\cos{\theta}_i\, d\phi_i$ are the connections of the two unit 2-spheres with metrics $ds^2_{S_i^2}$. The NS5-brane couples also to the  RR potentials, $C_1$ and $C_5$. In deriving these potentials from (\ref{F2}) and (\ref{F6}), we fix the integration constants such that,  for constant $\zeta$ with $0\leq\zeta\leq \pi/4$:
\begin{eqnarray}
\label{theC1}
C_1&=&-\frac{k}{2}\Bigl[ -\cos^2{\zeta}\, A_1+\sin^2{\zeta}\,A_2+\sin^2{\zeta}\,d\psi\Bigr]\\
C_5&=&\frac{1}{32}\,k\, L^4\sin^4{\zeta}\cos^2{\zeta}\sin{\theta}_1\sin{\theta}_2\, \, d\psi\wedge d\theta_1\wedge d\phi_1\wedge d\theta_2\wedge d\phi_2\, ,\nonumber
\end{eqnarray}
while, for constant $\zeta$ with $\pi/4 \leq\zeta\leq \pi/2$:
\begin{eqnarray}
\label{theC1bis}
C_1&=&-\frac{k}{2}\Bigl[ -\cos^2{\zeta}\, A_1+\sin^2{\zeta}\,A_2-\cos^2{\zeta}\,d\psi\Bigr]\\
C_5&=&-\frac{1}{32}\,k\, L^4\cos^4{\zeta}\sin^2{\zeta}\sin{\theta}_1\sin{\theta}_2\, \, d\psi\wedge d\theta_1\wedge d\phi_1\wedge d\theta_2\wedge d\phi_2\, . \nonumber
\end{eqnarray}
With these choices, the $S^2\times S^3$ structure of the metric is maintained and the RR 1-form potential, $C_{1}$, is proportional to the connection $A_{1}$ of the first $S^2$ parameterized by $(\theta_1, \phi_1)$ at $\zeta=0$, and to the connection $A_{2}$ of the second $S^2$ parameterized by $(\theta_2, \phi_2)$ at $\zeta=\pi/2$. Therefore the $N_{11}$ manifold can be continuously deformed into the first $S^2$ in the region $0\leq\zeta\leq \pi/4$ and into the second $S^2$ in the region $\pi/4 \leq\zeta\leq \pi/2$. Note that the two regions can be mapped onto each other by interchanging the two 2-spheres. The fact that the RR-potentials change sign under this exchange implies, however, that the NS5-branes in the $0\leq \zeta \leq \pi/4$ and $\pi/4\leq \zeta \leq \pi/2$ regions will have to be taken with opposite orientation in order to be stable.

\subsection{The action of the NS5-brane}

The metric and RR-fields which couple to the NS5-brane are isometric in the $\psi$ direction. Therefore the NS5-brane can be described by the action built in \cite{Herrero:2011bk}, suitable for the study of NS5-branes with $U(1)$ isometric worldvolume directions. We start by recalling the relevant features of this action. The reader is referred to \cite{Herrero:2011bk} for more details.

The (bosonic) worldvolume field content of the action built in \cite{Herrero:2011bk} consists on 4 transverse scalars, a vector field associated with D2-branes wrapping the isometric direction and ending on the NS5-brane, and a worldvolume scalar.  This scalar $c_{0}$ forms an invariant field strength $\mathcal{F}_{1}$ with the RR 1-form potential $C_{1}$ and is therefore associated with D0-branes.  Comparing with the action of the unwrapped NS5-brane \cite{BLO,EJL}, the self-dual 2-form field of the latter has been replaced by a vector $\mathcal{A}_{1}$ with $\mathcal{F}_{2} = d\mathcal{A}_{1}+\frac{1}{2\pi}P[i_l C_3]$. This is due to the fact that the D2-branes must now be wrapped on the isometric direction and hence their intersection with the NS5-brane becomes point-like under the dimensional reduction on $\psi$. This allows us to give a closed form for the action, since the self-duality condition on the 2-form field need not be imposed.  Note that the worldvolume dual of this vector field $\mathcal{A}_{1}$ is a 2-form $\tilde{\mathcal{A}}_{2}$ associated with D2-branes transverse to the isometric direction and ending on the NS5-brane.  This will be relevant for our discussion in the next subsection.

The BI part of the action is given by \cite{Herrero:2011bk}:
\begin{eqnarray}
\label{wrappedNS5}
&& \hspace{-1.0cm} S^{\mathrm{DBI}}_{\mathrm{NS5}}=-T_4 \int d^5\xi\,\, e^{-2\phi}\sqrt{l^2+e^{2\phi}(i_l C_1)^2} \, \times \nonumber\\
&& \hspace{-1.0cm} \hspace{1.37cm} \times \, \sqrt{ \left|\det{\left(P[{\cal G}]+\frac{e^{2\phi}\,l^2}{l^2+e^{2\phi}(i_l C_1)^2} \, {\cal F}_1^2
+\frac{2\pi\, e^\phi}{\sqrt{l^2+e^{2\phi}(i_l C_1)^2}} \, {\cal F}_2 \right)}\right|}.
\end{eqnarray}
Here $l^\mu=2\, \delta^{\mu}_{\psi}$ is an Abelian Killing vector\footnote{The factor of $2$ in this definition comes from the $0 \leq \psi \leq 4\pi$ range of the isometric coordinate.} that points along the isometric $U(1)$ direction, ${\cal G}$ is the reduced metric with components ${\cal G}_{\mu\nu}=g_{\mu\nu}-l^{-2}l_\mu l_\nu$ and $i_l C_p$ denotes the interior product of the $C_p$ potential with the Killing vector.  The field strength ${\cal F}_1=dc_0+P[C_1]$ is associated with D0-branes ``ending" on the NS5-brane. (The pull-back is taken with respect to the gauge covariant derivatives relative to the isometry,
${\cal D}_\xi X^\mu=\partial_\xi X^\mu-l^{-2}l_\nu\, \partial_\xi X^\nu l^\mu$.) In this way the dependence on the isometric direction is effectively eliminated.  In turn,
${\cal F}_2=d\mathcal{A}_1+\frac{1}{2\pi}P[i_l C_3]$ is the field strength associated with D2-branes wrapping the isometric direction $\psi$ and ending on the NS5-brane.  The action is therefore manifestly isometric under translations along the Killing direction.  Note that the vector field in ${\cal F}_2$, together with the 4 transverse scalars and the worldvolume scalar $c_0$, give the right counting of bosonic degrees of freedom, 8, in the effectively 5-dimensional worldvolume.
Finally, we have denoted by $T_4$ the tension of the brane to explicitly take into account that it is wrapped on the isometric direction. 

Finally, the relevant part of the CS action is given by \cite{Herrero:2011bk}:
\begin{equation}
\label{CSNS5}
S^{\mathrm{CS}}_{\mathrm{NS5}}=T_4 \int d^5\xi \,\, \Bigl\{P[i_l C_5] \wedge dc_0+ 2\pi P[C_3]\wedge {\cal F}_2\Bigr\}.
\end{equation}
The second term will vanish on our background, but it will be helpful to keep it in mind, nevertheless, when we later give an interpretation of the topological defects which must be dissolved on the worldvolume of the NS5-brane. Recall that the pull-backs are taken with respect to the gauge covariant derivatives associated with the isometry, which implies that $P[C_3]$ must be transverse to the isometric direction.

\subsection{Calculation of the energy}

Let us now substitute our ansatz for the NS5-brane, given in section 2.2, into the previous action. Let us start by first considering the region $0\leq \zeta \leq \pi/4$. 

 In order to make the brane stable we must induce D0-brane charge in the configuration. We do this by introducing a time dependent $c_0$. Its coupling to the RR-potential, $C_5$, in the CS action (\ref{CSNS5}) will then ensure stability. Moreover, we need to switch on a non-vanishing electric field proportional to the connection of the first two-sphere. The idea is to deform the worldvolume, as in \cite{Sadri:2003mx, Prokushkin:2004pv, AliAkbari:2007jr}, in such a way as to generate a BPS configuration. In this case, the electric field which produces the right deformation is given by
\begin{equation}
\label{electricfield}
\mathcal{F}_{2} =\frac{1}{8\pi}\, k L^{2} \,\sin{\zeta}\cos{\zeta}\, \,dt \wedge A_1 
= \frac{1}{8\pi}\, k L^{2} \,\sin{\zeta}\cos{\zeta}\, \cos{\theta_{1}} \, \,dt \wedge d\phi_{1} \, .
\end{equation}
One can check that this ansatz satisfies the equations of motion. This electric field induces a topological charge in the configuration:
\begin{equation}
\mathcal{J}_{3} \equiv d\mathcal{F}_{2} 
= \frac{1}{8\pi}\, k L^{2}\,\sin{\zeta}\cos{\zeta} \, \,dt \wedge \left( \sin{\theta_{1}} \, \, d\theta_{1} \wedge d\phi_{1} \right)  \neq 0
\hspace{0.25cm} \Longrightarrow \hspace{0.25cm} \mathcal{F}_{2} \neq d\mathcal{A}_{1}\, ,
\end{equation}
which is associated with fluctuations of 1-dimensional topological defects on the NS5-brane worldvolume.
Electric fields of this type play a role in the study of confining phases of $d$ dimensional $p$-form Abelian gauge theories, in the context of Mandelstam-'t Hooft duality \cite{Mandelstam:1976,tHooft:1978}. The idea in \cite{QT:1997} (see also \cite{Julia:1979ur}) is that, for a compact tensor field,  $A_p$, of rank $p$ in $d$-dimensions, a confined phase may arise after the condensation of $(d-p-1)$-dimensional topological defects. The fluctuations of a continuous distribution of topological defects generate new low-energy modes in the theory, which can be described by the non-exact part of the field strength $F_{p+1}$ in such a way that, away from the defects, $F_{p+1}=dA_{p}$ is exact. In our particular set-up, the non-exact part of ${\cal F}_2$ is associated with fluctuations of 1-dimensional topological defects, which we interpret as the intersections of open D2-branes, transverse to the isometric direction $\psi$, with the worldvolume of the NS5-brane.

Let us recall first that a D$p$-brane admits two types of topological defects: particles and $(p-3)$-branes. The first may be interpreted as the end-points of open strings and are therefore perturbative in origin. The second originate as the end-points of non-perturbative open D$(p-2)$-branes and can thus only be described by D$(p-2)$-brane degrees of freedom in the strong coupling regime. The above mechanism shows, however, that we can incorporate these degrees of freedom into the perturbative action. In the NS5-brane case, the perturbative fundamental strings are replaced by open D2-branes wrapping the isometric direction, while the non-perturbative ones are associated with D2-branes transverse to the isometric direction, both of which end on the NS5-brane.  This is inferred from the second coupling in (\ref{CSNS5}), which is analogous in the NS5-brane action to the coupling
\begin{equation}
S^{\mathrm{CS}}_{\text{D}p}=2\pi\, T_p \int P[C_{p-1}]\wedge F_2
\end{equation}
in the D$p$-brane action (with $F_2$ the BI field strength).

With these ingredients we can now calculate the DBI action of the NS5-brane in the region $0\leq \zeta \leq \pi/4$.  Substituting into (\ref{wrappedNS5}) the background given by (\ref{NS5ansatz}) and (\ref{theC1}), the electric field (\ref{electricfield}), and a time dependent $c_0(t)$, we arrive at
\begin{equation}
S^{\mathrm{DBI}}_{\mathrm{NS5}}= - \, kN \, \sin^3{\zeta}\cos^2{\zeta}\int dt\,\sqrt{1- \cos^2{\zeta}\,\left(\frac{2 \dot{c}_0}{k}\right)^2},
\end{equation}
where we have integrated over the two 2-spheres.
The CS action, in turn, reads
\begin{equation}
S^{\mathrm{CS}}_{\mathrm{NS5}}= kN \, \sin^4{\zeta}\cos^2{\zeta} \int dt\, \left(\frac{2\dot{c}_0}{k}\right)\, .
\end{equation}

Note that, both in the DBI and CS actions, $c_0$ is a cyclic coordinate and therefore its conjugate momentum, interpreted in this case as D0-brane charge, must be conserved. We denote it by $Q_{D0}$. The corresponding Hamiltonian is then a function of $\zeta$, the coordinate that parameterizes the varying radial size and the non-trivial fibre over the two 2-spheres. It reads
\begin{equation}
\label{energy}
H(\zeta)=\frac{k}{2}\, Q_{D0}\,\sqrt{1+\tan^2{\zeta}\,\Bigl(1-\frac{N}{2\,Q_{D0}}\sin^2{2\zeta}\Bigr)^2}\, .
\end{equation}
The ground state is then reached when $\zeta$=0 or when
\begin{equation}
\label{giantNS5}
\sin{2\zeta}=\sqrt{\frac{2\,Q_{D0}}{N}}\, 
\end{equation}
associated with $\dot{c}_{0} = k/2$.  For both solutions
\begin{equation}
E=\frac{k}{2}\, Q_{D0}\, ,
\end{equation}
i.e. the energy is that of $Q_{D0}$ D0-branes in the $AdS_4\times \mathbb{CP}^3$ background, each with an energy $k/2$ \cite{Aharony:2008ug}.
Note that the Hamiltonian (\ref{energy}) gives the energy of the NS5-brane in the region $0\leq \zeta \leq \pi/4$. Therefore we find two energetically degenerate minima.
For the first solution, $\zeta=0$, the NS5-brane shrinks to the first 2-sphere and can carry arbitrary units of D0-brane charge. It cannot, however, shrink to zero size. This happens because
the worldvolume geometry $S^{2} \times S^{3}$ is a topologically non-trivial manifold. For the solution given by (\ref{giantNS5}), the NS5-brane wraps a non-Einstein $N_{11}$ manifold whose $\lambda$, $\Lambda_1$, $\Lambda_2$ parameters depend on the D0-brane charge. Due to  (\ref{giantNS5}), this charge should satisfy $Q_{D0}\le N/2$.
The maximum D0-brane charge is reached when $\zeta=\pi/4$, for which the two 2-spheres have the same radius and the geometry of the $N_{11}$ manifold becomes
\begin{equation}
\label{extendedmetric}
ds^2_{\text{NS5}}= \frac{L^2}{4}\Bigl[ - \, dt^2 + (d\psi+A_1+A_2)^2+ds^2_{S^2_1}+ds^2_{S^2_2}\Bigl]\, .
\end{equation}
In this case, the energy is
\begin{equation}
E=\frac{kN}{4}
\end{equation}
and can be accounted for by either $N/2$ D0-branes or $k/2$ D4-branes dual to ABJM di-baryons. 
This is a manifestation of the symmetry of Young tableaux with $N$ rows and $k$ columns, as in \cite{Aharony:2008ug,Herrero:2011bk}.

Finally, the region $\pi/4\leq \zeta \leq \pi/2$ is described by interchanging the two 2-spheres in the above analysis. The background is given by (\ref{NS5ansatz}) and (\ref{theC1bis}).  An electric field must be turned on which is identical to (\ref{electricfield}), except that it involves the connection $A_{2}$ on the second 2-sphere.  The Hamiltonian as a function of $\zeta$ then takes the form
\begin{equation}
H(\zeta)=\frac{k}{2}\, Q_{D0}\,\sqrt{1+\cot^2{\zeta}\,\Bigl(1-\frac{N}{2\,Q_{D0}}\sin^2{2\zeta}\Bigr)^2}\, .
\end{equation}
Thus, in this case, there are again two energetically degenerate minima. For the first solution, $\zeta=\pi/2$, the NS5-brane shrinks to the second 2-sphere and can carry arbitrary units of D0-brane charge. For the second solution, given by (\ref{giantNS5}) with $\pi/4\leq \zeta \leq \pi/2$,
the NS5-brane wraps the same non-Einstein manifold defined by (\ref{extendedmetric}). 
\begin{figure}[htb!]
\begin{center} 
\hspace{-0.35cm}
\includegraphics[width=15.75cm,height=5.7cm]{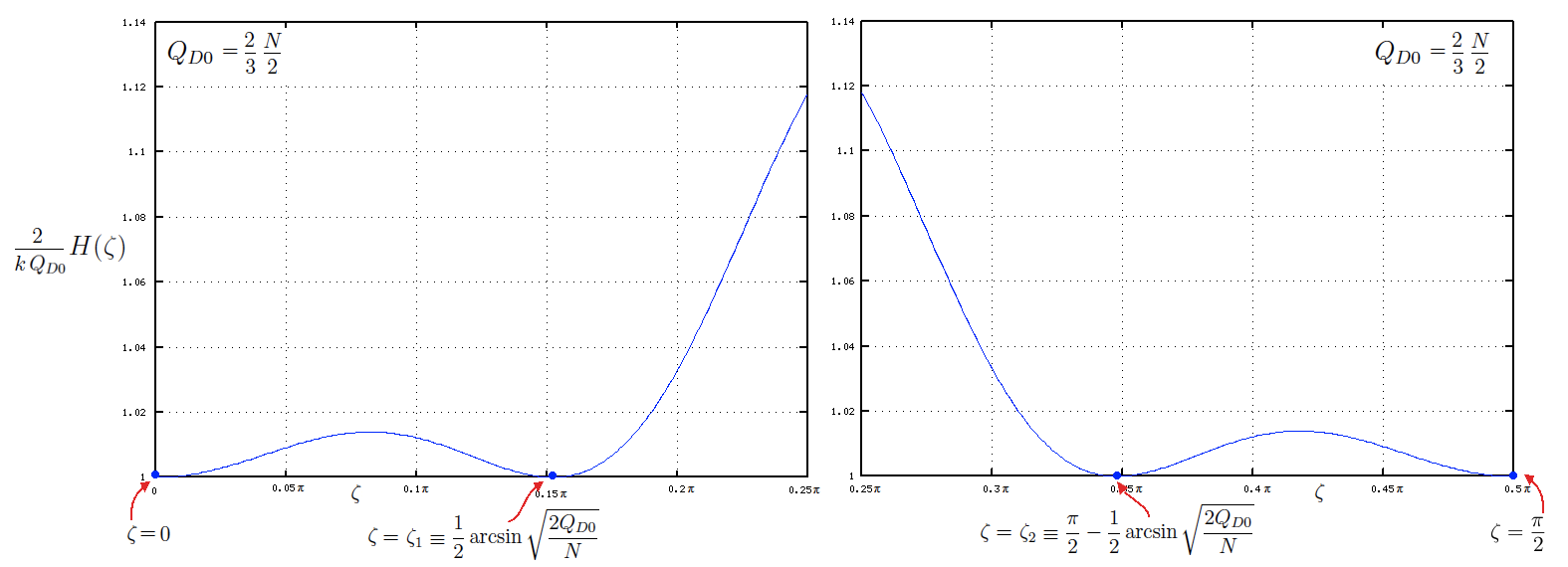}
\caption{The off-shell energy $H(\zeta)$ of the NS5-brane $S^{2}\times S^{3}$ configurations in units of $k \hspace{0.05cm} Q_{D0}/2$ as functions of the size $\zeta$ and
plotted at fixed D0-brane charge $Q_{D0} = N/3$ in the regions $0 \leq \zeta \leq \pi/4$ and $\pi/4 \leq \zeta \leq \pi/2$, respectively.}
\label{figure-energy}
\end{center}
\end{figure}
\vspace{-0.25cm}

Figure \ref{figure-energy} above shows the four energetically degenerate minima for $0\leq \zeta \leq \pi/2$.  There are two minima
\begin{equation} 
\label{minima}
\zeta_{1}=\frac12 \arcsin{ \sqrt{\frac{2Q_{D0}}{N}} } \hspace{0.75cm} \text{and} \hspace{0.75cm} \zeta_{2} =\frac{\pi}{2}-\frac12  \arcsin{ \sqrt{\frac{2Q_{D0}}{N}} } \, ,
\end{equation}
for which the topology of the expanded brane is $S^2\times S^3$, and two minima, at
$\zeta=0$ and $\zeta = \pi/2$, with $S^2$ topology. All of them are associated with BPS configurations. Although the expanded brane carries D0-brane charge it cannot shrink to a D0-brane due to its topologically non-trivial worldvolume.  When $Q_{D0}=N/2$, the two minima given by (\ref{minima}) coincide at $\zeta=\pi/4$.  Note that these maximal NS5-brane configurations with $Q_{D0}=N/2$ carry different electric fields and are therefore distinct degenerate BPS solutions.  They collapse onto different $S^{2}$'s contained in the worldvolume geometry.  This $N/2$ bound on the D0-brane charge is half the usual bound of $N$ on the D0-brane or angular momentum charge
(see, for example, \cite{McGreevy:2000cw,Grisaru:2000zn,Herrero:2011bk, Giovannoni:2011pn}).  We might speculate that each maximal NS5-brane is half of another maximal configuration, carrying $N$ units of charge, which collapses onto both $S^{2}$'s in the worldvolume geometry.

\begin{figure}[htb!]
\begin{center} 
\hspace{-0.35cm}
\includegraphics[width=15.75cm,height=5.5cm]{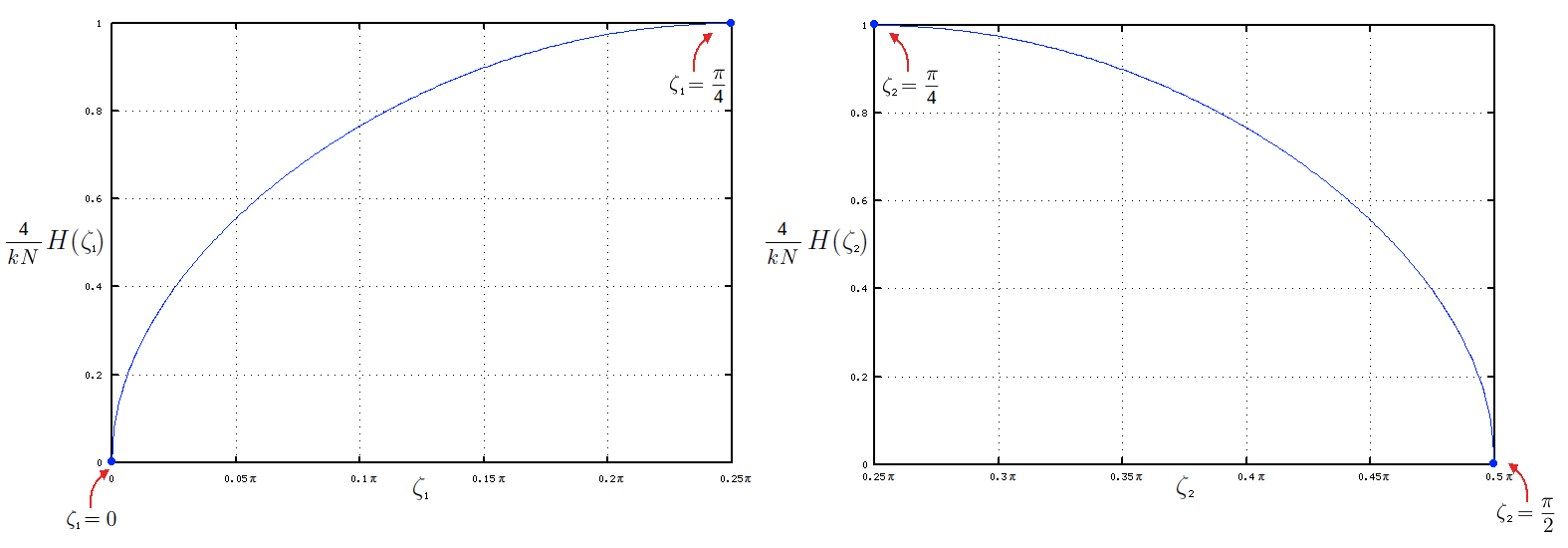}
\caption{The on-shell energies $H(\zeta_{1})$ and $H(\zeta_{2})$ of the NS5-brane $S^{2}\times S^{3}$ BPS solutions in units of $kN/4$, plotted as functions of $\zeta_{1}$ and $\zeta_{2}$, with  $0 \leq \zeta_{1} \leq \pi/4$ and 
$\pi/4 \leq \zeta_{2} \leq \pi/2$.}
\label{figure-energy2}
\end{center}
\end{figure}

The energies of the NS5-brane BPS $S^{2} \times S^{3}$ solutions are given by
\begin{equation}
E = \frac{kN}{4} \sin^{2}{2\zeta_{1}} \hspace{0.75cm} \text{and} \hspace{0.75cm} E = \frac{kN}{4} \sin^{2}{2\zeta_{2}}
\end{equation}
and are plotted as functions of the size parameters, $\zeta_{1}$ and $\zeta_{2}$, respectively, in figure \ref{figure-energy2}.  The D0-brane charge $Q_{D0}$ is not fixed here, but varies with the size parameter of the solution.
\vspace{-0.25cm}

The dual BPS operators in the ABJM model have conformal dimension $\Delta = k \, Q_{D0}/2$  and baryon number $k \, Q_{D0}$.  They are built from $k \, Q_{D0}$ bifundamental scalars\footnote{These scalars carry baryon number $1$ and conformal dimension $1/2$.} $A_{i}$ or $B^{\dag}_{i}$ with monopole operators attached to ensure gauge invariance, as in \cite{SheikhJabbari:2009kr,Berenstein:2009sa}.  The factor of $k$ is needed for these monopole operators to be equivalent to a large gauge transformation and hence invisible.  The monopole operators are constructed from a magnetic field in the ABJM conformal field theory and the magnetic charge is associated with the D0-brane charge $Q_{D0}$.  The interpretation in the ABJM model of the topological charge associated with 1-dimensional topological defects (D2-branes intersecting the NS5-brane) is, however, unclear.  The bound $Q_{D0} \leq N/2$ on the D0-brane charge dissolved on the NS5-brane is a realization of a stringy exclusion principle, similar to that of \cite{Maldacena:1998bw}.  There must exist an upper bound of $N/2$ on the magnetic charge in the ABJM field theory, which implies an upper bound of $kN/2$ on the $U(1)$ $\mathcal{R}$-charge and of $kN/4$ on the conformal dimension of the dual operators.

\section{A squashed $S^2\times S^3$ giant graviton}

In this section we show that the previous NS5-brane configuration, when lifted to eleven dimensions, takes the form of an M5-brane giant graviton which has the geometry of a squashed $S^2\times S^3$ of the $N_{10}$ type.

\subsection{The M5-brane ansatz}

Uplifting the ansatz for the NS5-brane in section 2.2 to eleven dimensions, we obtain an M5-brane with induced metric 
\begin{eqnarray}
\label{M5ansatz}
\hspace{-1.0cm} ds^2_{\text{M5}}&=& R^2\Bigl[- \, dt^2+\frac{1}{k^2} \, d\tau^2+\frac{1}{k}\,d\tau\, \Bigl(\cos^2{\zeta}\, A_1-\sin^2{\zeta}\, (d\psi+A_2)\Bigr)+ \nonumber\\
&& \hspace{0.68cm} + \, \frac14 \cos^2{\zeta}\,(ds^2_{S^2_1}+A_1^2)
+ \frac14 \sin^2{\zeta}\,\Bigl((d\psi+A_2)^2+ds^2_{S^2_2}\Bigr)\Bigr]\, ,
\end{eqnarray}
in the region $0\leq \zeta\leq \pi/4$, and
\begin{eqnarray}
\label{M5ansatzbis}
\hspace{-1.0cm} ds^2_{\text{M5}}&=& R^2\Bigl[- \, dt^2+\frac{1}{k^2} \, d\tau^2+\frac{1}{k}\,d\tau\, \Bigl(\cos^2{\zeta}\, (d\psi+A_1)-\sin^2{\zeta}\, A_2\Bigr)+\nonumber\\
&& \hspace{0.68cm} + \, \frac14 \cos^2{\zeta}\,\Bigl((d\psi+A_1)^2+ds^2_{S^2_1}\Bigr)
+ \frac14 \sin^2{\zeta}\,(ds^2_{S^2_2}+A_2^2)\Bigr]\, ,
\end{eqnarray}
for $\pi/4\leq \zeta\leq \pi/2$. Here $\tau$ is the eleventh direction and we have used that $L^2 k=R^3$, with $R$ the radius of the $S^7$ in eleven dimensions. These metrics for the M5-brane arise from the metric of the $AdS_4\times S^7/\mathbb{Z}_k$ background
\begin{equation}
ds^2= \frac{R^{2}}{4} \, ds^2_{AdS_4}+R^2\, \Bigl[\Bigl(\frac{1}{k} \, d\tau+\omega\Bigr)^2+ds^2_{\mathbb{CP}^3}\Bigr]\, ,
\end{equation}
with $r=0$ and $\zeta={\rm constant}$.  Here
\begin{equation}
\omega=\frac12 \cos^2{\zeta}\, A_1-\frac12\, \sin^{2}{\zeta} \, (d\psi+A_2)\, ,
\end{equation}
in the region $0\leq \zeta\leq \pi/4$, and  
\begin{equation}
\omega=\frac12 \cos^2{\zeta}\, (d\psi + A_1)-\frac12\, \sin^{2}{\zeta} \, A_2\, ,
\end{equation}
for $\pi/4\leq \zeta \leq \pi/2$. In both cases
\begin{equation}
d\omega=-\frac12 \sin{2\zeta}\, d\zeta \wedge (d\psi+A_1+A_2)+\frac12 \cos^2{\zeta}\, dA_1-\frac12 \sin^2{\zeta}\, dA_2\, .
\end{equation}
Note that the D0-brane charge of the NS5-brane configuration is translated into a momentum charge for the M5-brane giant graviton along the $\tau$ direction, so $\tau=\tau(t)$. The remaining spatial parts of the metrics describe $N_{10}$ manifolds of the type discussed in \cite{Candelas:1989js}, in which the $S^2$ is squashed.

Finally, we take the eleven dimensional 6-form potential at constant $\zeta$ to be 
\begin{equation}
\label{M5ansatz2}
C_6=-\frac{1}{32}\frac{R^6}{k}\sin^4{\zeta}\cos^2{\zeta}\sin{\theta_1}\sin{\theta_2}\,d\tau\wedge d\psi\wedge d\theta_1\wedge d\phi_1\wedge d\theta_2\wedge d\phi_2\, ,
\end{equation}
for $0\leq \zeta \leq \pi/4$, and
\begin{equation}
\label{M5ansatz2bis}
C_6=\frac{1}{32}\frac{R^6}{k}\cos^4{\zeta}\sin^2{\zeta}\sin{\theta_1}\sin{\theta_2}\,d\tau\wedge d\psi\wedge d\theta_1\wedge d\phi_1\wedge d\theta_2\wedge d\phi_2\, ,
\end{equation}
for $\pi/4\leq\zeta\leq \pi/2$. This again implies that M5-branes in the regions $0\leq \zeta \leq \pi/4$ and $\pi/4\leq \zeta \leq \pi/2$ have to be taken with opposite orientations to be stable.

\subsection{Calculation of the energy}

Since $\psi$ is again an isometric direction of both the metric and 6-form potential, $C_{6}$, which couple to the M5-brane, we can use the action for a wrapped M5-brane constructed in \cite{JLR} to describe it.  We start by briefly recalling some of the properties of this action. The reader is referred to \cite{JLR} for more details.

The (bosonic) worldvolume field content of the action built in \cite{JLR} consists on 5 transverse scalars and a vector field, associated with M2-branes wrapping the isometric direction and ending on the M5-brane.  As for the NS5-brane discussed in the previous section, the worldvolume 2-form field is substituted by a vector field and this allows for a closed form for the action. This is particularly relevant for the configuration that we describe in this section.
The relevant terms of the action constructed in \cite{JLR} read:
\begin{equation}
\label{theM5action}
S_{\text{M5}}=T_4 \int d^5\xi \, \, \left\{- l\, \sqrt{\left|{\rm det} \Bigl(P[{\cal G}] +\frac{2\pi}{l}{\cal F}_2\Bigr)\right|}+P[i_l C_6]\right\}.
\end{equation}
Here $l^\mu = 2 \, \delta^{\mu}_{\psi}$ is the Killing vector along the isometric direction $\psi$, with $l^2=g_{\mu\nu}l^\mu l^\nu$, $\cal G$ is the reduced metric defined in the previous section and $i_l C_p$ denotes the interior product of the $C_p$ potential with the Killing vector. The invariant field strength ${\cal F}_2=d\mathcal{A}_1+\frac{1}{2\pi}P[i_l C_3]$ is associated with M2-branes wrapping the isometric direction and ending on the M5-brane. When reducing along the isometric direction, this gives rise to the BI field strength of the D4-brane. 

Let us now substitute our ansatz for the M5-brane of the previous subsection into this action. 
We shall start by describing the region $0\leq \zeta\leq \pi/4$.
The coupling of the $C_6$ potential to the angular velocity of the brane $\dot{\tau}$ ensures stability. Further, we need to switch on a non-vanishing electric field proportional to the connection of the first two-sphere:
\begin{equation}
\label{elecflux}
\mathcal{F}_{2} = \frac{R^3}{8\pi} \sin{\zeta}\cos{\zeta}\, \, dt \wedge A_1
=  \frac{R^3}{8\pi} \, \sin{\zeta}\cos{\zeta} \cos{\theta_{1}} \, \, dt \wedge d\phi_{1} \, . 
\end{equation}
As for the NS5-brane of the previous section, this field induces a topological charge in the configuration, which is associated with one-dimensional solitonic line defects, interpreted in this case as the end-points of  open M2-branes transverse to the isometric direction.  As in the previous section, one can check that this ansatz satisfies the equations of motion. 

We arrive at the following BI action, after integration over the two 2-spheres:
\begin{equation}
S^{\text{DBI}}_{\text{M5}}= - \, kN \, \sin^3{\zeta}\cos^2{\zeta} \int dt\, 
\sqrt{1-\cos^2{\zeta}\,\left( \frac{2\dot{\tau}}{k}\right)^2}.
\end{equation}
The CS part in turn reads:
\begin{equation}
S^{\text{CS}}_{\text{M5}}= kN \, \sin^4{\zeta}\cos^2{\zeta} \int dt \,\left( \frac{2\dot{\tau}}{k} \right).
\end{equation}

Given that $\tau$ is a cyclic coordinate, its conjugate momentum, $P_\tau$, is conserved. The Hamiltonian reads
\begin{equation}
H(\zeta)=\frac{k}{2}\, P_\tau \,\sqrt{1+\tan^2{\zeta}\,\Bigl(1-\frac{N}{2P_\tau}\sin^2{2\zeta}\Bigr)^2}\, .
\end{equation}
Therefore, in the $0\leq \zeta \leq \pi/4$ region, there are two ground states, $\zeta=0$ and 
\begin{equation}
\label{giantgrav}
\sin{2\zeta}=\sqrt{\frac{2P_\tau}{N}}\, ,
\end{equation}
for which $\dot{\tau} = k/2$.  Both these BPS solutions have
\begin{equation}
E=\frac{k}{2}\, P_\tau\, 
\end{equation}
and are associated with giant gravitons propagating along the $\tau$ direction. For the first solution $P_\tau$ can be arbitrary, as for the point-like graviton solution in \cite{McGreevy:2000cw}. This solution is, however, not point-like, its geometry being that of a squashed $S^2$, with metric
\begin{equation}
ds^2=\frac{R^2}{4}\Bigl[ds^2_{S_1^2}+A_1^2\Bigr]\, .
\end{equation}
This happens because, as for the NS5-brane in the previous section, the M5-brane cannot shrink to zero size due to its topologically non-trivial $S^{2}\times S^{3}$ worldvolume.
The second solution, in turn, has a maximum angular momentum given by 
\begin{equation}
\label{bound}
P_\tau\le \frac{N}{2}
\end{equation} 
and corresponds to an M5-brane wrapped on the manifold
\begin{equation}
ds^2=\frac{R^2}{4}\Bigl[\cos^2{\zeta}\Bigl(ds^2_{S_1^2}+A_1^2\Bigr)+\sin^2{\zeta}\Bigl((d\psi+A_2)^2+ds^2_{S_2^2}\Bigr)\Bigr],
\end{equation}
with $\zeta$ depending on the angular momentum as in (\ref{giantgrav}). This corresponds to a $N_{10}$ geometry of the type discussed in \cite{Candelas:1989js} in which the $S^2$ is squashed. In particular, the configuration with maximum angular momentum corresponds to a $N_{10}$ 
in which the squashed $S^2$ and the $S^3$ have the same radii.

Finally, as in the previous section, the region $\pi/4\leq \zeta \leq \pi/2$ is described by interchanging the two 2-spheres. This gives rise to two energetically degenerate minima. For the first solution, $\zeta=\pi/2$, the M5-brane shrinks to the second 2-sphere and can carry arbitrary angular momentum. For the second solution, given by (\ref{giantgrav}) with $\pi/4\leq \zeta \leq \pi/2$,
the M5-brane wraps the same squashed $N_{10}$ manifold with the two 2-spheres interchanged, and carries an electric field proportional to the connection of the second 2-sphere and a momentum charge satisfying the bound (\ref{bound}).  Again, this bound on $P_{\tau}$ is a realization of a stringy exclusion principle, similar to that of \cite{Maldacena:1998bw}.

\section{Fuzzy $S^2\times S^3$ geometries}

In this section we show that the previous NS5 and M5-brane configurations have microscopic desciptions in terms of multiple D2 and M2-branes expanding into fuzzy 3-spheres ($S^{1}$ fibres over fuzzy 2-spheres) inside the $N_{11}$ or (squashed) $N_{10}$ manifolds. 
The expanded 5-branes then wrap a classical $S^2$, spanned by the 2-branes, times a fuzzy $S^3$. This provides a particular {\it fuzzification} of the $N_{11}$ and $N_{10}$ manifolds in which the $S^2$ submanifold is not made fuzzy. The obstruction for a complete {\it fuzzification} is the topologically non-trivial character of these manifolds, which prevents them arising from expanding point-like objects.

The interest of this microscopical description is that it allows one to explore the finite 't Hooft coupling region of the dual CFT. It is well known that the macroscopic description, in terms of a single expanded brane with lower dimensional charge dissolved on its worldvolume, and the microscopic description, in terms of multiple lower dimensional branes expanding into a fuzzy manifold, have complementary ranges of validity \cite{Myers:1999ps}. While the first is valid in the supergravity limit, the second is a good description when the mutual separation of the expanding branes is much smaller than the string length, so that they can be described by the $U(n)$ effective action constructed by Myers \cite{Myers:1999ps}. In the particular case of $n$ 2-branes expanding into a 3-dimensional manifold of radius $R$, the volume per brane goes like $R^3/n$, which is of order $l_s^3$. Hence the condition 
\begin{equation}
\label{regime}
R \sim n^{1/3} 
\end{equation}
sets the regime of validity of the microscopical description\footnote{Here we work in units in which $l_s = 1$.}. The macroscopical description is, in turn, valid for $R \gg 1$. Thus both descriptions are complementary for finite $n$, but should agree in the large $n$ limit, where they have common ranges of validity. In the backgrounds with CFT duals under consideration in this paper, the condition (\ref{regime}) implies for the 't Hooft parameter
\begin{equation}
\lambda \sim n^{4/3} \hspace{1.0cm} \text{and} \hspace{1.0cm} \lambda \sim n^2 \, 
\end{equation}
in the IIA string theory and M-theory, respectively.  Exploring the possibility of finite $n$ therefore allows us to explore the finite t'Hooft coupling $\lambda$ region of the dual CFT through the microscopic study of the corresponding dual brane system.

In our particular set-up, the microscopic description of the BPS $S^2\times S^3$ NS5 or M5-brane configurations may be useful in learning about the finite 't Hooft coupling behaviour of the corresponding dual CFT operators. In this section we provide a microscopic description in terms of D2 or M2-branes expanding into fuzzy versions of the $S^2\times S^3$ geometries. We show that the 2-branes can expand into these manifolds through suitable (dielectric) couplings to the background 5- and 6-form potentials. For D2-branes, however, we will see that, in order to find the right coupling to the 5-form potential, an action valid for the description of the strong coupling regime of the system must first be provided. This action will arise as the dimensional reduction of the action for coincident M2-branes with a transverse isometric direction constructed in  \cite{Lozano:2000aq}. As we will see, this action provides an alternative to the Myers action for coincident D2-branes, more suitable for an exploration of the strong coupling regime.

Given that our starting point for both the D2 and M2-brane microscopic descriptions is the action describing coincident M2-branes, we will reverse the order in the previous sections and start by exploring the microscopic description of the M5-brane giant graviton in terms of expanding M2-branes. For this purpose, we will start by reviewing the main features of the action constructed in \cite{Lozano:2000aq} and then see that it provides the right microscopic description of the M5-brane giant graviton constructed in section 3. Then, in section 4.2, we will reduce the M2-brane action to construct the right action for coincident D2-branes from which the NS5-brane BPS configuration of section 2 can be built. In both examples we will observe that the microscopic descriptions match the macroscopic calculations of the previous sections in the limit in which the number of 2-branes becomes large.

\subsection{The microscopical M5-brane giant graviton}
 
\subsubsection{The action for multiple M2-branes}
 
The action for multiple M2-branes in a space with an isometric transverse direction was constructed in \cite{Lozano:2000aq}. This action was obtained by uplifting the Myers action for multiple D2-branes to eleven dimensions. Although, in the non-Abelian case, the worldvolume duality transformation that allows us to map the BI vector field of a D2-brane onto the extra transverse scalar of the M2-brane cannot be made, we can still work with the uplifted action, which is, however, constrained to describe M2-branes in backgrounds with an isometric direction. 
As shown in \cite{Lozano:2000aq}, it is in fact due to the existence of this special isometric direction that a dielectric coupling to the eleven dimensional 6-form potential, $C_{6}$, can be constructed. This coupling will indeed be responsible for stabilizing the M2-branes in our particular system.

The BI part of the action constructed in  \cite{Lozano:2000aq} is given by:
\begin{equation}
\label{M2DBI}
S_{n \, \mathrm{M2}}^{\mathrm{DBI}}=-T_2 \int d^3\xi\, \, {\rm STr} \sqrt{|{\rm det}\bigl[E_{\alpha\beta}+E_{\alpha i}\, (Q^{-1}-\delta)^i_k\, E^{kj}E_{j\beta}]\, {\rm det}Q^i_j|}
\end{equation}
Here $E_{\alpha\beta}=(P[{\cal G}] + 2\pi \,{\cal F}_2)_{\alpha\beta}$ and $E_{ij} = (P[{\cal G}])_{ij}$, with $\cal G$ the reduced metric (defined in the previous sections), and ${\cal F}_2$ the field strength associated with M2-branes wrapping the isometric direction and ending on the M2-branes:
\begin{equation}
{\cal F}_2=d\mathcal{A}_1+\frac{1}{2\pi}P[i_l C_3].
\end{equation}
Note that ${\cal F}_2$ gives the BI field strength of the D2-branes when compactified on the isometric direction. Here $l^\mu$ is the Killing vector that points along this direction and $l^2=g_{\mu\nu} \, l^\mu l^\nu$.  The indices $\alpha,\beta$ are taken to run over the worldvolume directions and $i,j$ over the fuzzy transverse ones.  Here $Q$ is given by 
\begin{equation}
Q^i_j \equiv \delta^i_j+i\, \frac{l}{2\pi}\,[X^i, X^k] \,E_{kj}\, .
\end{equation}

The relevant terms in the CS action read:
\begin{equation}
\label{M2CS}
S_{n \, \mathrm{M2}}^{\mathrm{CS}}=T_2\, \int d^3\xi\, \, {\rm STr} \left\{\frac{i}{2\pi}\, P\Bigl[ (i_X i_X)\, i_l C_6\Bigr]
+2\pi P\left[\frac{l_1}{l^2}\right]\wedge {\cal F}_2\right\}.
\end{equation}
Here $l_1$ is the 1-form $l_1=l_\mu \, dx^\mu=g_{\mu\nu} \, l^\nu dx^\mu$ and $l_1/l^2$ is then identified with the momentum operator along the isometric direction (see \cite{Lozano:2000aq}). This term will be relevant below when we take our ansatz for the M2-branes.
Note that the dielectric coupling to $C_6$ can only be constructed due to its extra contraction with the Killing direction. This will be the coupling responsible for stabilizing the M2-branes in the expanded configuration.

\subsubsection{The M2-brane ansatz}

Let us start by discussing the region  $0\leq \zeta\leq \pi/4$.  We showed in section 3 that in this region the M5-brane giant graviton is degenerate with an M2-brane wrapping the first 2-sphere and propagating with arbitrary angular momentum on the orbifold direction, $\tau$. The natural ansatz for the microscopic description is therefore to consider M2-branes wrapping the first $S^2$, parameterized by $(\theta_1, \phi_1)$, and expanding into a fuzzy $S^3$.
Further, we need to switch on a worldvolume electric flux associated with M2-branes wrapping the isometric direction $\psi$ and ending on the system, since this was a key ingredient of the macroscopic description. As in that case, the electric field will provide the necessary deformation of the squashed 2-sphere to allow the formation of a BPS state. In this case the electric field is again given by (\ref{elecflux}), but is now associated with instantonic topological defects produced by gravitational waves ending on the M2-branes. This can be inferred from the second coupling in 
(\ref{M2CS}).

It is important to stress that, with the ansatz taken for the metric in the $0\leq \zeta\leq \pi/4$ region, given by (\ref{M5ansatz}), the M2-branes are stable and do not develop any tadpoles. This is not the case with the second ansatz (\ref{M5ansatzbis}) or with the more symmetric one (see for instance \cite{Nishioka:2008ib}):
\begin{equation}
\omega=\frac12 \cos^2{\zeta}\, A_1+\frac14\, (\cos^2{\zeta}-\sin^{2}{\zeta}) \, d\psi-\frac12 \, \sin^2{\zeta}\, A_2\, .
\end{equation}
Indeed, in these gauges, M2-branes wrapping the first $S^2$ would develop a tadpole through the coupling
\begin{equation}
\int_{\mathbb{R}\times S^2} P\left[\frac{l_1}{l^2}\right]\wedge {\cal F}_2\, ,
 \end{equation}
that would need to be cancelled by adding M2-branes wrapping the isometric direction and ending on them. Clearly this would not provide the right microscopic description, since the expanded M5-brane does not contain any tadpoles. 
This fact further justifies the reason why we need to take a different gauge for the metric in the two regions $0\leq \zeta\leq \pi/4$ and $\pi/4\leq \zeta\leq \pi/2$. Indeed, in the region $\pi/4\leq \zeta\leq  \pi/2$, with M2-branes wrapping the second 2-sphere parameterized by $(\theta_2,\phi_2)$ providing the right microscopic description, the metric (\ref{M5ansatzbis}) produces a vanishing tadpole on the worldvolume.

Let us now make the non-commutative ansatz. The structure of the metric (\ref{M5ansatz}) suggests that, in the region $0\leq \zeta\leq \pi/4$,
the M2-branes should expand into the fuzzy 3-sphere spanned by $(\psi,\theta_2,\phi_2)$. The fact that the action describing the M2-branes contains a transverse isometric direction, which is Abelian, further suggests that we should describe the fuzzy 3-sphere as an $S^1$ bundle, parameterized by $\psi$, over a fuzzy 2-sphere, parameterized by $(\theta_2,\phi_2)$. 
 Fuzzy 3-spheres realized as Abelian $S^1$ bundles over fuzzy $S^2$'s have been 
discussed in the literature in various contexts. They were first proposed as the concrete fuzzification of the 3-sphere giant graviton on $AdS_5\times S^5$ in \cite{Janssen:2003ri}.
More recently they have also been identified as the fuzzy manifolds on which M2-branes should expand to give an M5-brane in ABJM \cite{Nastase:2009ny}.

The description of the fuzzy 2-sphere is the standard one. Using Cartesian coordinates the condition
\begin{equation}
\sum_{i=1}^3 (x^i)^2=1
\end{equation}
can be imposed at the level of matrices by taking the $X^i$  in the totally symmetric irreducible 
representation of order $m$, with dimension $n=m+1$,
\begin{equation}
\label{noncom1}
X^i=\frac{1}{\sqrt{m(m+2)}}\, \, J^i
\end{equation}
with $J^i$ the generators of $SU(2)$. They then satisfy
\begin{equation}
\label{noncom}
[X^i,X^j]=i\frac{2}{\sqrt{m(m+2)}} \, \, \epsilon_{ijk} \, X^k\, .
\end{equation}

With these ingredients we can now calculate the energy of the expanding M2-branes using the action reviewed in the previous subsection. We will only discuss the $0\leq\zeta\leq\pi/4$ region. The region $\pi/4\leq\zeta\leq\pi/2$ is simply described by interchanging the role played by the two 2-spheres, while taking into account the different gauge for the metric and the 6-form potential.

\subsubsection{Calculation of the energy}

Substituting the background given by (\ref{M5ansatz}) and (\ref{M5ansatz2}), the electric field given by (\ref{electricfield}) and the non-commutative condition (\ref{noncom}) in the action (\ref{M2DBI}) and (\ref{M2CS}), we find, after integration over the 2-sphere parameterized by $(\theta_1,\phi_1)$,
\begin{eqnarray}
&& \hspace{-0.25cm} S_{M2} = -\,  kN \, \frac{m+1}{\sqrt{m(m+2)}} \times  \\
&& \hspace{-0.25cm} \hspace{0.84cm} \times 
 \int dt\, \left\{ \sin^3{\zeta}\cos^2{\zeta}\, \sqrt{1+\frac{m(m+2)}{2\,N\,k\,\sin^6{\zeta}}} \, \, \sqrt{1 - \cos^2{\zeta}\,\left( \frac{2\dot{\tau}}{k}\right)^2}\nonumber \, - \, \sin^4{\zeta}\cos^2{\zeta}\, \left(\frac{2\dot{\tau}}{k}\right) \right\},
\end{eqnarray}
where we have used that $E_{ij} = \tfrac{1}{4} \, R^{2} \sin^{2}{\zeta} \,\, \delta_{ij}$, from which it immediately follows that $E_{\alpha i}\, (Q^{-1}-\delta)^i_k\, E^{kj}E_{j\beta}=0$ in our background, and that hence
\begin{equation}
\label{thedet}
\det{Q}=\left(1+\frac{R^6 \sin^6{\zeta}}{16\,\pi^2\, m(m+2)}\right) \, \mathbb{I},
\end{equation}
which is exact in the limit
\begin{equation} \label{micro-limit}
R \gg 1\, , \hspace{0.5cm} n = m+1 \gg 1\, , \hspace{0.5cm} {\rm with} \hspace{0.5cm} \frac{R^3}{n} \propto \frac{\sqrt{N}}{n} = {\rm finite}  \qquad
\end{equation}  
(see section 5.1 of \cite{Lozano:2011dd} for a detailed discussion).  
Also, ${\rm STr}\, \mathbb{I}=(m+1)$.  It now becomes necessary to take $R^{3}/n$ to be a large finite number to approximate the new square root in the above action.

Transforming $\dot{\tau}$ to $P_\tau$, we obtain the Hamiltonian
\begin{equation}
H(\zeta)=\frac{k}{2} \, P_\tau \, \sqrt{1+\tan^2{\zeta}\,\Bigl(1-\frac{N}{2P_\tau}\frac{m+1}{\sqrt{m(m+2)}}\sin^2{2\zeta}\Bigr)^2}\, .
\end{equation}
Therefore, in the region $0 \leq \zeta \leq \pi/4$, we find again two ground states, $\zeta=0$ and
\begin{equation}
\label{giantgrav2}
\sin^2{2\zeta}={\frac{2P_\tau}{N}}\frac{\sqrt{m(m+2)}}{m+1} = {\frac{2P_\tau}{N}}\frac{\sqrt{(m+1)^2-1}}{m+1}, 
\end{equation}
both with energy 
\begin{equation}
E=\frac{k}{2} \, P_\tau\, .
\end{equation}
While for the first solution $P_\tau$ can be arbitrary, for the second solution it must satisfy
\begin{equation}
P_\tau\le \frac{N}{2}\frac{m+1}{\sqrt{(m+1)^{2}-1}} = \frac{N}{2}\frac{n}{\sqrt{n^{2}-1}}.
\end{equation}
Comparing the Hamiltonian, radius and upper bound with their counterparts in the M5-brane macroscopic derivation, we find that, as expected, there is exact agreement in the large $n=m+1$ limit.  Note that this BPS solution is only valid in the  large $\lambda/n^2 \propto R^{6}/n^2$ regime.  This calculation should therefore be seen as a first step towards exploring the finite t'Hooft coupling region, as well as a verification of the macroscopic computation at the level of the microscopic description.

\subsection{The microscopical NS5-brane solution}

In this subsection we show that a very similar microscopic description exists for the NS5-brane with D0-brane charge of section 2 in terms of expanding D2-branes. Indeed, the type IIA description of $n$ M2-branes expanding into the M5-brane giant graviton which we have just discussed should be in terms of $n$ D2-branes expanding into an NS5-brane with D0-brane charge. However, as we mentioned, there is no dielectric coupling in the Myers action for coincident D2-branes which can possibly explain the expansion of multiple D2-branes into an NS5-brane. The resolution of this puzzle is that we need to use an action to describe this system of D2-branes which is more suitable for the study of the strong coupling regime. Indeed, the configuration in which D2-branes expand into an NS5-brane is related by T-duality to one in which D3-branes expand into an NS5-brane in type IIB, which is, in turn, S-dual to a configuration in which D3-branes expand into a D5-brane. The expansion of the D2-branes into an NS5-brane is therefore intrinsically non-perturbative. This is also evident from the $e^{-2\phi}$ coupling in front of the action for the NS5-brane. 

We show next that a non-perturbative action for coincident D2-branes can be constructed by simply reducing the action for coincident M2-branes of the previous subsection along a transverse direction different from the isometric direction. Note that the Myers action is recovered when the reduction is performed along the isometric direction. Indeed, interchanging the two compact directions produces the desired mapping onto the strong coupling regime.

\subsubsection{The action for multiple D2-branes}

As we have just mentioned, the action suitable for the study of D2-branes expanding into an NS5-brane is built by reducing the action for M2-branes discussed in the previous subsection.  This action will still contain the isometric direction $\psi$, which will allow the construction of the right dielectric coupling responsible for the expansion of the D2-branes into an NS5-brane.

Let us start by discussing the CS part of the action. The reduction of the CS action (\ref{M2CS}) to type IIA produces the couplings\footnote{There are some extra couplings which will not be relevant to our discussion.}
\begin{equation}
\label{C5}
S_{n \, \mathrm{D2}}^{\mathrm{CS}}=i \, \frac{T_2}{2\pi} \int d^3\xi \, \,  {\rm STr}\left\{P\Bigl[(i_X i_X)\, i_l C_5\Bigr] \wedge dc_0
+2\pi\,P\left[\frac{l_1+e^{2\phi}C_1 (i_l C_1)}{l^2+e^{2\phi} (i_l C_1)^2}\right]\wedge {\cal F}_2 \right\}.
\end{equation}
In the first term $C_5$ is contracted with the Killing direction and it is coupled to the worldvolume scalar $c_0$ associated with D0-branes, which we introduced in our discussion of the NS5-brane. This term will indeed be responsible for the expansion of the D2-branes into an NS5-brane. The second term will be used, in turn, to justify the absence of tadpoles.

Reducing the DBI action (\ref{M2DBI}) we obtain
\begin{equation}
S^{\mathrm{DBI}}_{n \, \textrm{D2}}=-T_2\int d^3\xi \,\, {\rm STr}\left\{e^{-\phi} \, \sqrt{\left|{\rm det}
\left[E_{\alpha\beta}+\frac{l^2 \, e^{2\phi}}{l^2 + e^{2\phi}(i_l C_1)^2} \, ({\cal F}_1)_{\alpha\beta}^2\right] \,
\det{Q^i_j} \right|} \,\, \right\},
\end{equation}
where we have taken for simplicity $E_{\alpha i}\, (Q^{-1}-\delta)^i_k \, E^{kj}E_{j\beta}=0$, since this will be the case in our background.
In this action 
\begin{equation}
E_{\alpha\beta}=(P[{\cal G}])_{\alpha\beta}+\frac{2\pi\, e^\phi}{\sqrt{l^2+e^{2\phi}(i_l C_1)^2}} \,\,({\cal F}_2)_{\alpha\beta},
\end{equation} 
with ${\cal G}$ the reduced metric, and ${\cal F}_1=dc_0 +P[C_1]$ and ${\cal F}_2=d\mathcal{A}_1+\frac{1}{2\pi}P[i_l C_3]$ the field strengths associated with  
D0 and D2-branes, respectively, wrapping the isometric direction $\psi$ and ending on the multiple coincident D2-branes.
As expected for a strong coupling description, the dynamics of the branes is induced by D-branes ending on them and not by fundamental strings. Finally, $Q$ is given by
\begin{equation}
Q^i_j=\delta^i_j+i\, \frac{e^{-\phi}}{2\pi}\sqrt{l^2+e^{2\phi}(i_l C_1)^2}\, [X^i,X^k] \Bigl({\cal G}_{kj}+\frac{l^2 \, e^{2\phi}}{l^2+e^{2\phi}(i_l C_1)^2}(C_1-\frac{l_1}{l^2}\, i_l C_1)^2_{kj}\Bigr),
\end{equation}  
with $E_{ij} = (P[{\cal G}])_{ij}$ the reduced metric of the three fuzzy transverse directions. Let us now use this action to compute the energy of the D2-brane system in the background described in section 2.2.

\subsubsection{Calculation of the energy}

We shall first make an ansatz for our set of coincident D2-branes, which will be very similar to the ansatz for the coincident M2-branes taken in the previous subsection.  Again, we just describe the $0\leq \zeta \leq \pi/4$ region. The description in the region $\pi/4\leq \zeta \leq \pi/2$ is obtained simply by making the changes explained in section 2. We take D2-branes wrapping the 2-sphere parameterized by $(\theta_1, \phi_1)$ in the background defined by (\ref{NS5ansatz}) and (\ref{theC1}). It is important to stress that, with this choice, the D2-branes do not develop any tadpole through the coupling
\begin{equation}
\int_{\mathbb{R}\times S^2} P\left[\frac{l_1+e^{2\phi}C_1 (i_l C_1)}{l^2+e^{2\phi} (i_l C_1)^2}\right]\wedge {\cal F}_2
\end{equation}
in the CS action (\ref{C5}). To account for the D0-brane charge of the expanded NS5-brane, we switch on a time dependent worldvolume scalar field $c_0(t)$. Further, we induce instantonic topological defects (produced, in this case, by gravitational waves ending on the system) through a non-vanishing worldvolume electric field, given by (\ref{electricfield}), associated with D2-branes wrapping $\psi$ and ending on the system. The non-commutative ansatz for the transverse coordinates $X^{i}$ is the same as the one taken in the previous section.

With these ingredients, we finally arrive at a Hamiltonian
\begin{equation}
H(\zeta)=\frac{k}{2}\, Q_{D0}\, \sqrt{1+\tan^2{\zeta}\,\Bigl(1-\frac{N}{2\,Q_{D0}}\frac{m+1}{\sqrt{m(m+2)}}
\sin^2{2\zeta}\Bigr)^2},
\end{equation}
where $Q_{D0}$ accounts for the D0-brane charge. Here again we have a limit similar to (\ref{micro-limit}) and we take $kL^2/n$ to be finite, but large. There are two ground states, at $\zeta=0$ and
\begin{equation}
\sin^2{2\zeta}={\frac{2Q_{D0}}{N}}\frac{\sqrt{m(m+2)}}{m+1}, 
\end{equation}
in this region $0 \leq \zeta \leq \pi/4$, both with energy 
\begin{equation}
E=\frac{k}{2} \, P_\tau\, .
\end{equation}
While for the first solution $P_\tau$ can be arbitrary, for the second solution it must satisfy
\begin{equation}
Q_{D0}\le \frac{N}{2}\frac{m+1}{\sqrt{m(m+2)}} = \frac{N}{2}\frac{n}{\sqrt{n^{2}-1}}.
\end{equation}
Comparing the Hamiltonian, radius and upper bound with their counterparts in the macroscopical NS5-brane derivation, we again find exact agreement in the large $n=m+1$ limit.

\section{Conclusions}

In this paper we have constructed new BPS 5-brane configurations  with $S^2\times S^3$ topologies in the gravity dual of the ABJM model. One such configuration is an NS5-brane with D0-brane charge wrapping an $N_{11}$ non-Einstein manifold. This NS5-brane is degenerate in the ground state with a D2-brane with D0-brane charge $Q_{D0}$ wrapping the $S^2$ submanifold of the $N_{11}$. While this spherical D2-brane can have arbitrary D0-brane charge, the expanded NS5-brane must satisfy the bound $Q_{D0}\leq N/2$. 

This bound is very similar to the one encountered in giant graviton configurations. Indeed, lifted to eleven dimensions, this configuration is described by an M5-brane with angular momentum $P_{\tau}$ wrapping an $N_{10}$ manifold. In this case, the ground state is degenerate with an M2-brane with angular momentum wrapping the (squashed) $S^2$ submanifold of the $N_{10}$. While the spherical M2-brane can have arbitrary momentum, the expanded M5-brane must satisfy $P_\tau\leq N/2$. 

Although our giant graviton configuration satisfies what we might refer to as the defining properties of  a giant graviton, it differs from the standard D3-brane giant gravitons on $AdS_{5}\times S^{5}$
 \cite{McGreevy:2000cw,Grisaru:2000zn}, and from the D2 and D4-brane giant gravitons on $AdS_{4}\times \mathbb{CP}^{3}$ of \cite{Nishioka:2008ib,Giovannoni:2011pn} in three main features:
First, the expanded configuration is not degenerate with a point-like graviton with arbitrary momentum, but with an extended object. This is due to the topologically non-trivial character of the $S^2\times S^3$ manifold on which it is wrapped, which prevents its collapse to zero size.  Second, a topological charge $\mathcal{J}_{3} = d\mathcal{F}_{2}$ must be introduced in order for the energy to satisfy the BPS bound $E = k \hspace{0.05cm} P_{\tau}/2$.  This additional charge is non-perturbative in nature and is associated with 1-dimensional topological defects, interpreted as M2-branes ending on the M5-brane giant graviton.  Third, the bound imposed by the geometry on the angular momentum is not the rank of the gauge group $N$, but rather $N/2$.  The maximal M5 giant graviton, carrying $N/2$ units of momentum charge, occurs at $\zeta = \pi/4$ and collapses onto only one 2-sphere in the worldvolume geometry. One possible reason for the unusual N/2 bound might be that our maximal giant graviton is essentially half of another maximal configuration, carrying $N$ units of momentum charge, which collapses onto both 2-spheres.  Constructing this configuration explicitly would require, however, a unified treatment of the two regions in which the 2-branes wrapping the two 2-spheres can be defined simultaneously.

We have also seen that, microscopically, these 5-brane configurations with topologically non-trivial $S^2\times S^3$ geometries are explained in terms of $n$ D2 or M2-branes wrapping the $S^2$ submanifold and expanding, due to the Myers dielectric effect, into a fuzzy version of the $S^3$ (an $S^{1}$ bundle over a fuzzy $S^{2}$).  This microscopic description should be seen as a first step towards exploring the finite t'Hooft coupling region of the gauge theory.
In this set-up, the maximal size configuration is again reached when $\zeta=\pi/4$, for which the D0-brane or momentum charge acquires its maximum value.

An interesting line of future research is the construction of the BPS operators in the ABJM model which are dual to these 5-brane $S^{2}\times S^{3}$ configurations.  These operators will be built from the bifundamental scalars  $A_{i}$ or $B_{i}^{\dag}$ (associated with microscopic descriptions in terms of 2-branes wrapping the first or second 2-spheres) with monopole operators attached in such a way as to ensure gauge invariance, as discussed in \cite{SheikhJabbari:2009kr,Berenstein:2009sa}.  The upper bound of $N/2$ on the D0-brane charge and angular momentum in the eleventh direction indicates an upper bound of $N/2$ on the magnetic charge in the ABJM field theory.  This, in turn, implies an upper bound of $kN/2$ on the $U(1)$ $\mathcal{R}$-charge of the dual BPS operators.  In this way, the bound of $N/2$ on the D0-brane and angular momentum charges must be a realization of a stringy exclusion principle, similar to that of \cite{Maldacena:1998bw}.  The dual interpretation of the topological charge associated with 1-dimensional solitonic line defects (2-branes intersecting the 5-brane) remains unclear. It would be most interesting to clarify this point, especially since it is likely to provide insight into the non-trivial topology of this 5-brane $S^{2}\times S^{3}$ configuration from the perspective of the ABJM model.

An issue which we have not explored in this paper is the amount of supersymmetry preserved by our 5-brane configurations, beyond verifying that they satisfy BPS bounds.  It is known that the type IIA ten-dimensional and eleven-dimensional supergravity backgrounds $AdS_{4}\times \mathbb{CP}^{3}$ and $AdS_{4}\times S^{7}/\mathbb{Z}_{k}$ preserve 24 of a maximal 32 supersymmetries (unless $k = 1, 2$ for which there is supersymmetry enhancement) 
\cite{Grisaru:2000zn,Nishioka:2008ib}.   In computing the kappa-symmetry conditions, it will be necessary to work with a reduced action and take into account the field strength $\mathcal{F}_{2}$ associated with 2-branes ending on the 5-brane.  We leave this as an open problem for future analysis.

\subsection*{Acknowledgements}	
We would like to thank Jeff Murugan for useful discussions.
Y. L. is partially supported by the research grants FPA2012-35043-C02-02 and MEC-DGI-CSD2007-00042.


\begin{thebibliography}{99}

\bibitem{Maldacena:1997re}
  J.~M.~Maldacena, ``{\it The Large N limit of superconformal field theories and supergravity},''
  Adv.\ Theor.\ Math.\ Phys.\  {\bf 2} (1998) 231
  \texttt{[arXiv:hep-th/9711200]}.

\bibitem{McGreevy:2000cw}
  J.~McGreevy, L.~Susskind and N.~Toumbas, ``{\it Invasion of the giant gravitons from Anti-de Sitter space},''
  J. High Energy Phys. {\bf 0006} (2000) 008
  \texttt{[arXiv:hep-th/0003075]}.
  
\bibitem{Grisaru:2000zn}
  M.~T.~Grisaru, R.~C.~Myers and O.~Tafjord, ``{\it SUSY and goliath},''
  J. High Energy Phys. {\bf 0008} (2000) 040
  \texttt{[arXiv:hep-th/0008015]}.

\bibitem{Balasubramanian:2001nh}
  V.~Balasubramanian, M.~Berkooz, A.~Naqvi and M.~J.~Strassler, ``{\it Giant gravitons in conformal field theory},''
  J. High Energy Phys. {\bf 0204} (2002) 034
  \texttt{[arXiv:hep-th/0107119]}.

\bibitem{Corley:2001zk}
  S.~Corley, A.~Jevicki and S.~Ramgoolam,
  ``{\it Exact correlators of giant gravitons from dual N=4 SYM theory},''
  Adv.\ Theor.\ Math.\ Phys.\  {\bf 5} (2002) 809
  \texttt{[arXiv:hep-th/0111222]}.

\bibitem{de Mello Koch:2007uu}
  R.~de Mello Koch, J.~Smolic and M.~Smolic, ``{\it Giant Gravitons - with Strings Attached (I)},''
  J. High Energy Phys. {\bf 0706} (2007) 074
  \texttt{[arXiv:hep-th/0701066]}.

\bibitem{de Mello Koch:2007uv}
  R.~de Mello Koch, J.~Smolic and M.~Smolic,
  ``{\it Giant Gravitons - with Strings Attached (II)},''
  J. High Energy Phys. {\bf 0709} (2007) 049
  \texttt{[arXiv:hep-th/0701067]}.

\bibitem{Bekker:2007ea}
  D.~Bekker, R.~de Mello Koch and M.~Stephanou, ``{\it Giant Gravitons - with Strings Attached (III)},''
  J. High Energy Phys. {\bf 0802} (2008) 029
  \texttt{[arXiv:0710.5372]}.

\bibitem{Balasubramanian:2004nb}
  V.~Balasubramanian, D.~Berenstein, B.~Feng and M.~X.~Huang, ``{\it D-branes in Yang-Mills theory and emergent gauge symmetry},''
  J. High Energy Phys. {\bf 0503} (2005) 006
  \texttt{[arXiv:hep-th/0411205]}.

\bibitem{Berenstein:2006qk}
  D.~Berenstein, D.~H.~Correa and S.~E.~Vazquez,
  ``{\it A Study of open strings ending on giant gravitons, spin chains and integrability},''
  J. High Energy Phys. {\bf 0609} (2006) 065
  \texttt{[arXiv:hep-th/0604123]}.

\bibitem{Maldacena:1998bw}
  J.~M.~Maldacena and A.~Strominger,
  ``{\it $AdS_3$ black holes and a stringy exclusion principle},''
  J. High Energy Phys. {\bf 9812} (1998) 005
  \texttt{[arXiv:hep-th/9804085]}.

\bibitem{Berenstein:2005aa}
  D.~Berenstein,
  ``{\it Large N BPS states and emergent quantum gravity},''
  J. High Energy Phys. {\bf 0601} (2006) 125
  \texttt{[arXiv:hep-th/0507203]}.

\bibitem{Berenstein:2006yy}
  D.~Berenstein and R.~Cotta,
  ``{\it Aspects of emergent geometry in the AdS/CFT context},''
  Phys.\ Rev.\ D {\bf 74} (2006) 026006
  \texttt{[arXiv:hep-th/0605220]}.

	
\bibitem{LLM}
H.~Lin, O.~Lunin and J.~M.~Maldacena,
``{\it Bubbling AdS space and 1/2 BPS geometries},''
J. High Energy Phys. {\bf 0410} (2004) 025, \texttt{[arXiv:hep-th/0409174]}. 

\bibitem{Aharony:2008ug}
  O.~Aharony, O.~Bergman, D.~L.~Jafferis and J.~Maldacena,
  ``{\it N=6 superconformal Chern-Simons-matter theories, M2-branes and their gravity duals},''
  J. High Energy Phys. {\bf 0810} (2008) 091
  \texttt{[arXiv:0806.1218]}.

\bibitem{Mikhailov}
A. Mikhailov, ``{\it Giant gravitons from holomorphic surfaces}'', J. High Energy Phys. {\bf 0011} (2000) 027 \texttt{[arXiv:hep-th/0010206]}.
  
\bibitem{Berenstein:2008dc}
  D.~Berenstein and D.~Trancanelli,
  ``{\it Three-dimensional $\mathcal{N}=6$ SCFT's and their membrane dynamics},''
  Phys.\ Rev.\ D {\bf 78} (2008) 106009
  \texttt{[arXiv:0808.2503]}.

\bibitem{SheikhJabbari:2009kr}
  M.~M.~Sheikh-Jabbari and J.~Simon,
  ``{\it On Half-BPS States of the ABJM Theory},''
  J. High Energy Phys. {\bf 0908} (2009) 073
  \texttt{[arXiv:0904.4605]}.

\bibitem{Berenstein:2009sa}
  D.~Berenstein and J.~Park,
  ``{\it The BPS spectrum of monopole operators in ABJM: Towards a field theory description of the giant torus},''
  J. High Energy Phys. {\bf 1006} (2010) 073
  \texttt{[arXiv:0906.3817]}.

\bibitem{Dey}
T. Dey, ``{\it Exact Large $R$-charge Correlators in ABJM Theory}'', 
J. High Energy Phys. {\bf 1108} (2011) 066 
\texttt{[arXiv:1105.0218]}.
  
\bibitem{dMMMP}
R.~de Mello Koch, B.~A.~E.~Mohammed, J.~Murugan and A.~Prinsloo, ``{\it Beyond the Planar Limit in ABJM}"
J. High Energy Phys, {\bf 1205} (2012) 037,
\texttt{[arXiv:1202.4925]}

\bibitem{Caputa:2012}
P.~Caputa and B.~A.~E.~Mohammed, ``{\it From Schurs to Giants in ABJ(M)}"
J. High Energy Phys. {\bf 1301} (2013) 055 \texttt{[arXiv:1210.7705]}. 
   
\bibitem{Nishioka:2008ib}
  T.~Nishioka and T.~Takayanagi, ``{\it Fuzzy Ring from M2-brane Giant Torus},''
  J. High Energy Phys. {\bf 0810} (2008) 082
  \texttt{[arXiv:0808.2691]}.

\bibitem{Herrero:2011bk}
  M.~Herrero, Y.~Lozano and M.~Picos,
  ``{\it Dielectric 5-Branes and Giant Gravitons in ABJM},''
  J. High Energy Phys. {\bf 1108} (2011) 132
  \texttt{[arXiv:1107.5475]}.

\bibitem{Giovannoni:2011pn}
  D.~Giovannoni, J.~Murugan and A.~Prinsloo,
  ``{\it The Giant graviton on $AdS_{4} \times \mathbb{CP}^{3}$ - another step towards the emergence of geometry},''
  J. High Energy Phys. {\bf 1112} (2011) 003
  \texttt{[arXiv:1108.3084]}.

\bibitem{Hamilton:2009iv}
  A.~Hamilton, J.~Murugan, A.~Prinsloo and M.~Strydom,
  ``{\it A Note on dual giant gravitons in $AdS_{4} \times \mathbb{CP}^3$},''
  J. High Energy Phys. {\bf 0904} (2009) 132
  \texttt{[arXiv:0901.0009]}.

\bibitem{Kim}
J.~Kim, N.~Kim and J.~H.~Lee, ``{\it Rotating Membranes in $AdS_4 \times M^{1,1,1}$},"
J. High Energy Phys. {\bf 1003} (2010) 122 
\texttt{[arXiv:1001.2902]}.

\bibitem{Carballo:2009ei}
  J.~L.~Carballo, A.~R.~Lugo and J.~G.~Russo,
  ``{\it Tensionless supersymmetric M2 branes in $AdS_4 \times S^7$ and Giant Diabolo},''
  J. High Energy Phys. {\bf 0911} (2009) 118
  \texttt{[arXiv:0909.4269]}.

\bibitem{GLR}
N.~Guti\'{e}rrez, Y.~Lozano and D.~Rodr\'{i}guez-G\'{o}mez, ``{\it Charged particle-like branes in ABJM}'', 
J. High Energy Phys. \textbf{1009} (2010) 101 
\texttt{[arXiv:1004.2826]}.

\bibitem{Murugan:2011zd}
  J.~Murugan and A.~Prinsloo,
  ``{\it ABJM Dibaryon Spectroscopy},''
  J. High Energy Phys. {\bf 1105} (2011) 129
  \texttt{[arXiv:1103.1163]}.

\bibitem{Hirano:2012vz}
  S.~Hirano, C.~Kristjansen and D.~Young,
  ``{\it Giant Gravitons on $AdS_4 \times \mathbb{CP}^3$ and their Holographic Three-point Functions},''
  J. High Energy Phys. {\bf 1207} (2012) 006
  \texttt{[arXiv:1205.1959]}.

\bibitem{Candelas:1989js}
  P.~Candelas and X.~C.~de la Ossa, ``{\it Comments on Conifolds},''
  Nucl.\ Phys.\ B {\bf 342} (1990) 246.

\bibitem{Myers:1999ps}
  R.~C.~Myers, ``{\it Dielectric branes},''
  J. High Energy Phys. {\bf 9912} (1999) 022
  \texttt{[arXiv:hep-th/9910053]}.
 
\bibitem{Janssen:2003ri}
  B.~Janssen, Y.~Lozano and D.~Rodr\'{i}guez-G\'{o}mez,
  ``{\it A Microscopical description of giant gravitons. 2. The $AdS_{5} \times S^5$ background},''
  Nucl.\ Phys.\ B {\bf 669} (2003) 363
  \texttt{[arXiv:hep-th/0303183]}.

\bibitem{Nastase:2009ny}
  H.~Nastase, C.~Papageorgakis and S.~Ramgoolam, ``{\it The Fuzzy $S^2$ structure of M2-M5 systems in ABJM membrane theories},''
  J. High Energy Phys. {\bf 0905} (2009) 123
  \texttt{[arXiv:0903.3966]}.

 
\bibitem{Nilsson:1984bj}
  B.~E.~W.~Nilsson and C.~N.~Pope,
  ``{\it Hopf Fibration Of Eleven-dimensional Supergravity},''
  Class.\ Quant.\ Grav.\  {\bf 1} (1984) 499.

\bibitem{Hamilton:2010sv}
  A.~Hamilton, J.~Murugan and A.~Prinsloo,
  ``{\it Lessons from giant gravitons on $AdS_{5}\times T^{1,1}$},''
  J. High Energy Phys. {\bf 1006} (2010) 017
  \texttt{[arXiv:1001.2306]}.

\bibitem{BLO}
E.~Bergshoeff, Y.~Lozano, T.~Ort\'{\i}n, ``{\it Massive branes}" Nucl. Phys. {\bf B518} (1998) 363 
\texttt{[arXiv:hep-th/9712115]}.

\bibitem{EJL}
E.~Eyras, B.~Janssen, Y.~Lozano, ``{\it  Five-branes, K K monopoles and T duality,}" Nucl. Phys. {\bf B531} (1998) 275 
\texttt{[arXiv:hep-th/9806169]}.

\bibitem{Sadri:2003mx}
  D.~Sadri and M.~M.~Sheikh-Jabbari,
  ``{\it Giant hedgehogs: Spikes on giant gravitons,}''
  Nucl.\ Phys.\ B {\bf 687} (2004) 161
  \texttt{[arXiv:hep-th/0312155]}.

\bibitem{Prokushkin:2004pv}
  S.~Prokushkin and M.~M.~Sheikh-Jabbari,
  ``{\it Squashed giants: Bound states of giant gravitons,}''
  J. High Energy Phys. {\bf 0407} (2004) 077
  \texttt{[arXiv:hep-th/0406053]}.
  
\bibitem{AliAkbari:2007jr}
  M.~Ali-Akbari and M.~M.~Sheikh-Jabbari,
 ``{\it Electrified BPS Giants: BPS configurations on Giant Gravitons with Static Electric Field,}''
  J. High Energy Phys. {\bf 0710} (2007) 043
  \texttt{[arXiv:0708.2058]}.

\bibitem{Mandelstam:1976}
S.~ Mandelstam, ``{\it Vortices and quark confinement in non-Abelian gauge theories,}'' Phys. Rep. {\bf C23} (1976) 237.

\bibitem{tHooft:1978}
G.~'t Hooft, {\it On the phase transition towards permanent quark confinement,}'' Nucl. Phys. {\bf B138} (1978) 1.

\bibitem{QT:1997}
F. Quevedo, C. Trugenberger, ``{\it Phases of Antisymmetric Tensor Field Theories}," Nucl. Phys. {\bf B501} (1997) 143  \texttt{[arXiv:hep-th/9604196]}.

\bibitem{Julia:1979ur}
  B.~Julia and G.~Toulouse,
 ``{\it The Many Defect Problem: Gauge Like Variables For Ordered Media Containing Defects,}''
J. Physique Lett. {\bf 40} (1979) 396.

\bibitem{JLR}
B. Janssen, Y. Lozano, D. Rodr\'{\i}guez-G\'omez, ``{\it Giant gravitons and fuzzy $\mathbb{CP}^2$,}" Nucl. Phys. {\bf B712} (2005) 371 
\texttt{[arXiv:hep-th/0411181]}.

\bibitem{Lozano:2000aq}
  Y.~Lozano,
  ``{\it Noncommutative branes from M-theory},''
  Phys.\ Rev.\ {\bf D64} (2001) 106011
  \texttt{[arXiv:hep-th/0012137]}.


\bibitem{Lozano:2011dd}
  Y.~Lozano, M.~Picos, K.~Sfetsos and K.~Siampos,
  ``{\it ABJM Baryon Stability and Myers effect},''
  J. High Energy Phys. {\bf 1107} (2011) 032
  \texttt{[arXiv:1105.0939]}.



\end{thebibliography}
\end{document}